\documentclass[%
 reprint,
superscriptaddress,
notitlepage,
superscriptaddress,
nofootinbib,
 amsmath,amssymb,
 aps,
prd, 
floatfix,
]{revtex4-2}
\usepackage{graphicx}
\usepackage{dcolumn}
\usepackage{bm}
\usepackage{epsfig}
\usepackage{amsmath}
\input{epsf}
\usepackage{psfrag}
\usepackage{xcolor}
\usepackage{ulem}

\newcommand{\bea}{\begin{eqnarray}}
\newcommand{\eea}{\end{eqnarray}}
\newcommand{\nn}{\nonumber}

\begin{document}

\title{Bjorken $x$ weighted Energy-Energy Correlators from the Target Fragmentation Region to the Current Fragmentation Region}

\author{Haotian Cao}
\email{haotiancao@mail.bnu.edu.cn}
 \affiliation{Department of Physics, Beijing Normal University, Beijing, 100875, China}
\author{Hai Tao Li}
\email{haitao.li@sdu.edu.cn}
 \affiliation{School of Physics, Shandong University, Jinan, Shandong 250100, China}
\author{Zihao Mi}
 \affiliation{Department of Physics, Beijing Normal University, Beijing, 100875, China}

\begin{abstract}
We present the complete spectrum for the Bjorken $x$ weighted Energy-Energy Correlation in the deep inelastic scattering (DIS) process, from the target fragmentation region to the current fragmentation region, in the Breit frame. The corresponding collinear and transverse momentum-dependent logarithms are resummed to all orders with the accuracy of NLL and  N$^3$LL, respectively. And the results in the full region are matched with ${\cal O}(\alpha^2_s)$ fixed-order calculation. The final numerical predictions are presented for both EIC and CEBAF kinematics.
\end{abstract}

\maketitle

\section{Introduction}

The pursuit of a comprehensive understanding of the nucleonic structure and the intricate mechanisms underlying the formation of hadrons from partons and beam remnants is a paramount objective in the field of particle physics. 
This quest will continue to be at the forefront of scientific exploration within the Standard Model, particularly at the forthcoming electron-ion collider (EIC) and future QCD facilities~\cite{AbdulKhalek:2021gbh,DPAPreport2022,Accardi:2023chb}.

In recent years, approaches to nucleon/nucleus tomography have significantly evolved and enriched, since many novel ideas have been proposed in the field. One notable advancement is the jet-based studies of the transverse momentum dependent (TMD) structure functions~\cite{Gutierrez-Reyes:2018qez,Liu:2018trl,Gutierrez-Reyes:2019msa,Gutierrez-Reyes:2019vbx,Arratia:2020nxw,Liu:2020dct,Arratia:2020ssx,Li:2020rqj,Kang:2020fka,H1:2021wkz,Kang:2021kpt,Liu:2021ewb,Kang:2021ffh,Li:2021gjw,Lai:2022aly,Kang:2022dpx,Arratia:2022oxd} and the study of gluon saturation~\cite{vanHameren:2014lna,Hatta:2016dxp,Liu:2022ijp,Caucal:2022ulg,Wang:2022zdu,Ganguli:2023joy,Caucal:2023nci,vanHameren:2023oiq,Caucal:2023fsf}.

Event shape observables, such as thrust and C-parameter, serve as measures of the energy flow, multiple particle correlations, and radiative patterns in high-energy collisions, which have undergone extensive investigations at various colliders, and have played a central role in enhancing our understanding of the perturbative and non-perturbative dynamics of QCD over the past several decades.

Energy-Energy Correlation (EEC)~\cite{Basham:1978bw,Basham:1978zq,Hofman:2008ar,Belitsky:2013ofa,Belitsky:2013xxa,Kologlu:2019mfz,Korchemsky:2019nzm,Dixon:2019uzg,Chen:2020vvp,ALI1984447,Gao:2019ojf}
is an event shape originally introduced in the context of $e^+ e^-$ collisions as an alternative to the thrust family. EEC stands out among other event shape observables for its simplicity and effectiveness in revealing the intrinsic transverse-dependent dynamics~\cite{Li:2020bub,Ali:2020ksn,Li:2021txc} and the scales of the quark-gluon plasma~\cite{Yang:2023dwc,Andres:2022ovj,Andres:2023xwr,Andres:2023ymw}. It has also been used to study cold nuclear matter effects~\cite{Devereaux:2023vjz}.
Moreover, owing to the high perturbative accuracy achieved both in resummed and fixed-order calculation~\cite{Moult:2018jzp,Ebert:2020sfi,Gao:2019ojf,Li:2020bub}, complemented by high precision measurements~\cite{Abe_1994,ADEVA1991469,AKRAWY1990159,DECAMP1991479,ATLAS:2015yaa,ATLAS:2017qir,ATLAS:2020mee}, EEC offers opportunities for precision studies in QCD. In particular, EEC has been used for precise extractions of the strong coupling constant, illuminating the effects of the intrinsic mass of the elementary particles of QCD~\cite{craft2022beautiful,Holguin:2023bjf} and understanding gluon saturation and nuclear modifications~\cite{Kang:2023oqj}. Meanwhile, 
instead of using calorimetry, 
track-based measurements can be utilized~\cite{Chen:2020vvp,Li:2021zcf,Jaarsma:2023ell,Jaarsma:2022kdd,Lee:2023xzv,Lee:2023npz,Lee:2023tkr}, providing high pointing and angular resolution. 

In Ref.~\cite{Li:2021txc}, the EEC has been adapted to the deep inelastic scattering (DIS) process in the current fragmentation region (CFR) of the Breit frame. It was shown that the EEC in this region can be used to extract the conventional TMD parton distribution functions (TMDPDFs) and the TMD fragmentation functions (TMDFFs). The associated TMD resummation was carried out at the N$^3$LL level of accuracy. 

On the other hand, in the target fragmentation region (TFR), where the outgoing particles propagate in the forward direction close to the incoming hadron beam,  
a variant of EEC, named nucleon energy-energy correlator (NEEC) was proposed in~\cite{Liu:2022wop}, which supplies a unique opportunity to reveal the intrinsic dynamics of nucleons. Notably, similar to EEC, NEEC manifests a remarkable phase transition between the perturbative parton and the non-perturbative free hadron phase~\cite{Liu:2022wop}. NEEC has also been shown powerful in unraveling the on-set of gluon saturation~\cite{Liu:2023aqb} predicted by small-$x$ physics. Furthermore, a joint measurement of NEEC in the TFR and CFR exhibits an exquisite signature of the linearly polarized gluons inside the nucleons~\cite{Li:2023gkh}. The derivation of the NEEC factorization theorem and its NLL resummation were obtained in~\cite{Cao:2023oef}.

\begin{figure}[htbp]
  \begin{center}
   \includegraphics[scale=0.7]{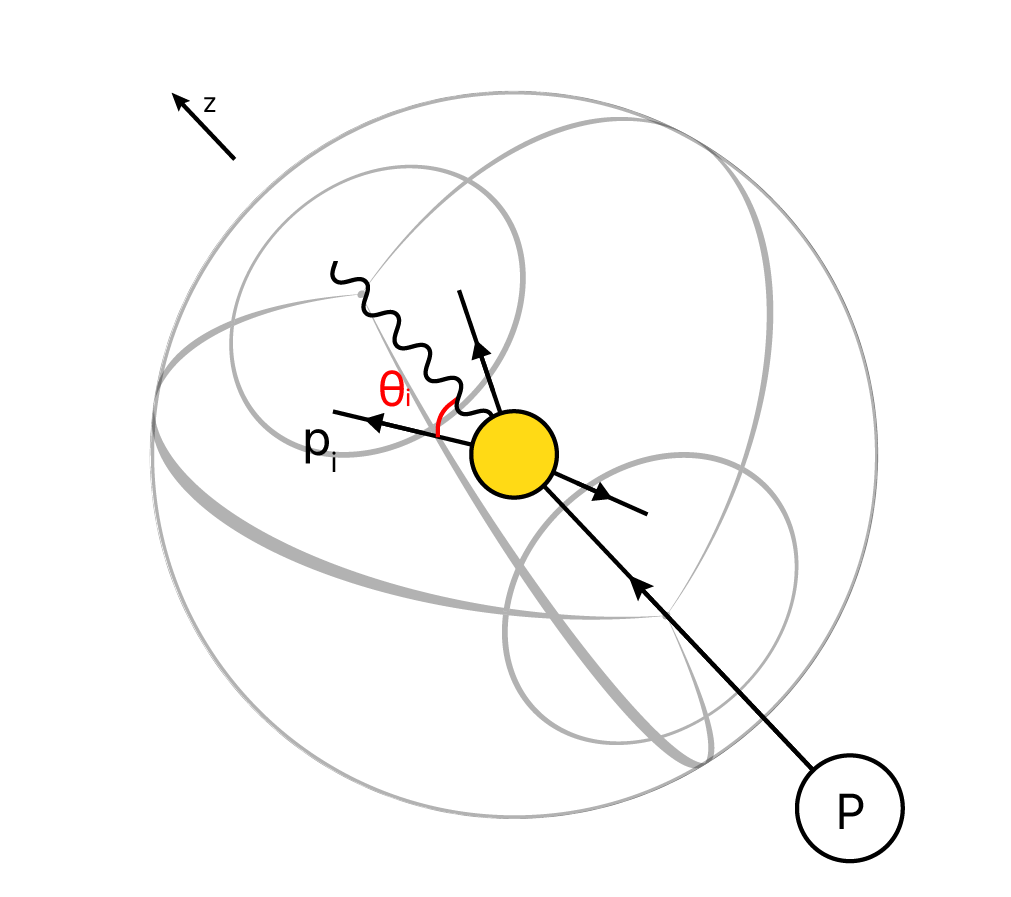} 
\caption{The EEC measurement in DIS.}
\label{fg:TFRCFR}
 \end{center}
\end{figure}

In this paper, we investigate the Bjorken $x$ weighted EEC in TFR and CFR region, which was first introduced by~\cite{Liu:2022wop} and called NEEC in their paper. The definition is 
\bea\label{eq:eec-def1} 
\frac{d\Sigma_N}{dQ^2d\theta} = \sum_i \int d\sigma(x_B,Q^2,p_i) x_B^{N-1} \frac{E_i}{E_P}
\delta(\theta - \theta_i) \,. 
\eea
Here $N > 1$ is a positive power, and $d\sigma$ is the differential cross section with Bjorken $x_B$ and virtuality of the photon $Q$. $p_i$ denotes the four-momentum of the particle detected by the calorimetry. The angle $\theta_i$ is the polar angle of $p_i$ with respect to the nucleon beam.  $E_i$ and $E_P$ are the energy of the detected particle and the incoming nucleon, respectively. The sphere represents the detector that reports the energy and the angle of the final state particle. The measurement is illustrated in Fig.\ref{fg:TFRCFR}. In the rest of the paper, we will use EEC to represent the Bjorken $x$ weighted EEC defined in Eq.~(\ref{eq:eec-def1}) for abbreviation.


In the TFR and TMD region, EEC can be systematically analyzed using the factorized formula based on the soft-collinear effective theory (SCET)~\cite{Bauer:2000ew,Bauer:2000yr,Bauer:2001ct,Bauer:2001yt,Bauer:2002nz,Beneke:2002ph}. We perform a consistency check of our approach by comparing the LO and NLO singular distributions, obtained by the factorization formula. We show the resummed cross section in the TMD region up to N$^3$LL level of accuracy and in the TFR up to NLL level of accuracy. The resummed distribution was matched with the NLO fixed-order QCD result for both EIC and CEBAF kinematics. The non-perturbative effects are discussed briefly in the EIC and CEBAF.

The paper is organized as follows. In Section \ref{sec:obs} we introduce the definitions of EEC and the kinematics formula in both the TFR and the TMD region. In Section \ref{sec:fac} we introduce the factorized formula in both the TFR and the TMD region. In Section \ref{sec:num} we present the numerical calculations of the resummation and the fixed-order singular distribution and compare our results with {\tt PYTHIA} simulations. We conclude in Section \ref{sec:con}. 

\section{Kinematics}
\label{sec:obs}
In this paper, we examine the process of DIS, where $k^\mu, k'^\mu$, and $P^\mu$ represent the four-momenta of the initial electron, the outgoing electron, and the initial-state proton, respectively. The momentum of the virtual photon is given by $q\equiv k -k'$.
The Lorentz invariant variables are conventionally defined as follows:
\bea
Q^2 \equiv -q^2, \qquad x_B \equiv \frac{Q^2}{2P\cdot q}, \qquad z_i\equiv\frac{P\cdot p_i }{P\cdot\sum p_i}\, ,
\eea 
where $Q^2$ represents the virtual photon momentum squared, $x_B$ is the Bjorken scaling variable, and $z_i$ denotes the momentum fraction carried by the observed particles ($p_i$) with respect to the sum of all observed particles.
In our analysis, we work in the Breit frame, where a distinct separation between the target and current fragmentation region can be defined by the hemispheres that cover the $+z$ and $-z$ directions, respectively. The momentum of the virtual photon only acquires momentum in its $z$ component:
\bea
q^{\mu}=\frac{Q}{2}(\bar{n}^\mu-n^\mu)=Q(0,0,0,-1)\,,
\eea 
with $\bar{n}^\mu\equiv(1,0,0,-1)$ and $n\equiv(1,0,0,1)$. The proton carries the momentum.
\bea
P^{\mu}=\frac{Q}{2x_B}n^\mu=\frac{Q}{2x_B}(1,0,0,1)\,.
\eea 
Throughout this paper, we adopt the standard notation, where $p^+ \equiv \bar{n}\cdot p$ and $p^- \equiv n\cdot p$. Here, $n\equiv(1,0,0,1)$ and $\bar{n}\equiv(1,0,0,-1)$ in the light-cone basis, with a vector denoted as $p^\mu = (p^+, p^-,\bold{p}_T)$. Consequently, in the Breit frame, $z_i=\frac{p_i^-}{Q}$.

We can relate the vector in any frame where the momentum of the virtual photon is $(q^+,q^-,\bold{q}_T)$ to the Breit frame with $v_{B}^\mu=(\Lambda R)^{\mu\nu} v_{\text{any},\nu}$, where 
\begin{equation}
    R^{\mu\nu}\equiv 
        \begin{pmatrix}
            1&0&0&0\\
            0&\frac{q_1}{q_T}&\frac{q_2}{q_T}&0\\
            0&\frac{-q_2}{q_T}&\frac{q_1}{q_T}&0\\
            0&0&0&1
        \end{pmatrix}
\end{equation}

\begin{equation}
    \Lambda^{\mu\nu}\equiv 
        \begin{pmatrix}
            \frac{q_0}{Q}+\frac{Q}{q^-}&-\frac{q_T}{Q}&0&-\frac{q_3}{Q}-\frac{Q}{q^-}\\
            -\frac{q_T}{Q}&1&0&\frac{q_T}{q^-}\\
            0&0&1&0\\
            \frac{q_0}{Q}&-\frac{q_T}{Q}&0&-\frac{q_3}{Q}
        \end{pmatrix}
\end{equation}
The angle of the final state particle $i$ can be defined as $\arctan(p_{i,T}/p_{i,3})$.

The kinematic region can be roughly divided into two distinct parts: the current fragmentation region (CFR), the target fragmentation region (TFR)~\cite{Mulders:2000jt,Boglione:2016bph,Gonzalez-Hernandez:2018ipj,Boglione:2019nwk,Boglione:2022gpv}. Each region has its own characteristics and is associated with different aspects of the scattering process. 

In the CFR, the observed particles result from the fragmentation of the parton struck by the virtual photon. The outgoing parton fragments into the detected particles. The CFR can be further divided into two subregions. The first subregion is the TMD region, where $\theta-\pi\ll 1$. In this region, the momenta of the observed particles scale as $p_i\sim Q(\theta^2,1,\theta)$. TMD factorization theorems are well-established and applicable in this case. The second subregion is the hard region, where $\theta\sim 1$, and the momenta of the observed particles scale as $p_i\sim Q(1,1,1)$. In this region, hard QCD radiations produce a large hadronic transverse momentum, and it is appropriate to deal with the fixed-order QCD calculations based on collinear factorization theorems.

In TFR, $\theta\ll 1$, the momenta of the observed particles scale as $p_i\sim Q(1,\theta^2,\theta)$. The TFR is associated with the fragmentation of spectator partons, which originate in the target nucleon but do not experience a hard collision with the virtual photon. These partons continue to move predominantly in the direction of the parent nucleon, where the measured hadron predominantly travels in the forward direction of the incoming target. 

In the hard region, the distribution is very well described by the fixed-order QCD calculations, while in the TFR and TMD region, resummation of enhanced logarithms is required for reliable predictions. To this end, the cross-section can be factorized, with the framework of SCET.
\begin{figure}[htbp]
  \begin{center}
   \includegraphics[scale=1.0]{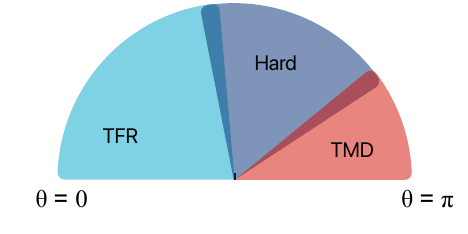} 
  \caption{Sketch of kinematical regions of EEC in terms of the Breit frame angle. }
  \label{fg:TFRCFR2}
  
 \end{center}
\end{figure}
Events in both regions can be used to comprehend the internal structure of hadrons and the properties of strong interactions. 
To better visualize the different kinematic regimes discussed above, we provide an angle map in Fig.\ref{fg:TFRCFR2}. 

\section{Bjorken $x$ weighted EEC SPECTRUM}

\label{sec:fac}

\subsection{TMD region}
\label{subsec:BR}
In this section, we briefly review the factorization of the EEC in the TMD region where $\pi-\theta\ll 1$. The factorization is similar to the one in \cite{Li:2021txc} except for the factor of the weighted Bjorken $x_B^{N-1}$. 

In the TMD region, the EEC can be related to the single hadron production process $e+p\rightarrow e+a+X$ with a small transverse momentum of the observed hadron. The expression for $\Sigma_N(Q^2,\theta)$ is given by:
\bea
    \frac{d\Sigma_N}{dQ^2d\theta}&=&\int dx_B x_B^{N-1}\sum_a\int d^2\bold{q}_Tdz
    \nn \\        
    &\times&
    \frac{d\sigma_{e+p\rightarrow e+a+X}}{dQ^2dx_Bd^2\bold{q}_Tdz}\frac{E_a}{E_p}\delta(\theta_{ap}-\theta)\,.
\eea 
The TMD cross section can be expressed in terms of TMD PDF and FF as follows:
\bea
    \frac{d\sigma_{e+p\rightarrow e+a+X}}{dQ^2dx_Bd^2\bold{q}_Tdz}&=&
    H(Q^2,\mu;\mu_H)
        \int  d^2{\bold q}_{T}\,  \frac{d^2\bold{b}}{(2\pi)^2} 
        \nn \\
         &&\hspace{-10.ex}\times\exp[-i{\bold q}_T\cdot \bold{b}]B_{f/p}\left(b,x_B,\mu,\nu;\mu_B,\nu_B\right)\nn \\
        && \hspace{-10.ex}\times S(b,\mu,\nu;\mu_S,\nu_S)
       D_{a/f}(z,b,\mu,\nu;\mu_J,\nu_J)\,,
\eea 
where $B_{f/p}$ is the TMD beam function, $S$ is the soft function, $D_{a/f}$ is the fragmentation function for parton $f$ to hadron $a$, and $b=|\mathbf{b}|$. The factorization of the EEC can be obtained by approximating $\frac{E_a}{E_p}$ as $x_Bz_a$:
\bea
\label{eq:TMDfac}
    \frac{d\Sigma_N}{dQ^2 d\theta}&=&\int dx_B\, x_B^{N}
        H(Q^2,\mu;\mu_H)
        \int  d^2{\bold q}_{T}\,  \frac{d^2\bold{b}}{(2\pi)^2} 
        \nn \\
        &\times& \exp[-i{\bold q}_T\cdot \bold{b}]B_{f/p}\left(b,x_B,\mu,\nu;\mu_B,\nu_B\right)\nn \\
         &\times&S(b,\mu,\nu;\mu_S,\nu_S)
       J_{f,\text{EEC}}(b,\mu,\nu;\mu_J,\nu_J)\nn \\
         &\times&     
        \delta\left(\frac{2|{\bold q}_{T}|}{Q}-\theta\right)\, ,
\eea 
where $J_{f,\text{EEC}}$ is the EEC (anti-)quark jet function defined as the first moments of the fragmentation functions:
\bea
    J_{f,\text{EEC}}\equiv\sum_a\int_0^1 dz\, zD_{f/a}(z,\bold{b})\,.
\eea 
When $1\gg(\pi-\theta) Q \gg \Lambda_{\rm QCD}$, through an OPE, the EEC jet function and TMD beam function can be expressed in terms of a convolution of short distance matching coefficients.
\bea
\label{eq:EECjetope}
    &&J^{\text{OPE}}_{f,\text{EEC}}=\sum_j\int_0^1 d\omega \, \omega{\cal D}_{fj}\left(\frac{b}{\omega},\omega,\mu,\nu\right)\, ,
\\
&&B^{\text{OPE}}_{f/p}(b, x,\mu,\nu;\mu_B,\nu_B) =\sum_{i} \int_x^1 \frac{dz}{z} 
\nn \\
&&\,\times {\cal I}_{fi}\left(b, \frac{x}{z},\mu,\nu;\mu_B,\nu_B\right) f_{i/p}(z,\mu)\, ,
\eea
where ${\cal D}_{fj}$ and ${\cal I}_{fi}$ are the matching coefficients, and $f_{i/p}$ represents the parton distribution functions. We have used the superscript OPE to denote that this is the leading contribution in the expansion and is considered a good approximation of the true TMDs in the perturbative regime where $(\pi-\theta)Q\gg \Lambda_{\text{QCD}}$.

The beam, jet, and soft functions can be evolved to the common scale $\mu$ from their natural scales at $\nu_B$, $\nu_J$, $\nu_S$, $\mu_B$, $\mu_J,$ and $\mu_S$, respectively, as
\bea
\label{eq:BJSevolPos}
B_f(x,b,\mu,\nu;\mu_B,\nu_B) &=& U_B(\mu,\nu;\mu_B,\nu_B)
\nn \\
&&\times
 B_f(x,b,\mu_B,\nu_B), \nn \\
J_{f,\rm{EEC}}(b,\mu,\nu;\mu_J,\nu_J) &=& U_J(\mu,\nu;\mu_J,\nu_J) \nn \\
&&\times J_{f,\rm{EEC}}(b,\mu_J,\nu_J),  \\
S_{EEC}(b,\mu,\nu;\mu_S,\nu_S) &=& U_S(\mu,\nu;\mu_S,\nu_S)\nn \\
&&\times S_{EEC}(b, \mu_S,\nu_S),\nn
\eea
where $U_B$, $U_J$, and $U_S$ are the position space evolution factors for the beam, jet, and soft functions, respectively. Similarly, the hard function also has a multiplicative renormalization group evolution
\bea
\label{eq:hardfuncevol}
H(Q^2, \mu; \mu_H) &=& U_H (Q^2,\mu, \mu_H)H(Q^2, \mu_H)\,,
\eea
where $U_H (\xi^2,\mu, \mu_H)$ is the corresponding hard function renormalization group evolution factor.

The RG-evolved cross-section reads
\bea
\label{eq:TMDfacRG}
    \frac{d\Sigma_N}{dQ^2d\theta}&=&\int dx_B\, x_B^{N}
        H(Q^2,\mu_H)
        \int  d^2{\bold q}_{T}\,  \frac{d^2\bold{b}}{(2\pi)^2} 
        \nn \\
        &&\hspace{-10.ex}\times \exp[-i{\bold q}_T\cdot \bold{b}]U_{tot}B_{f/p}\left(b,x_B,\mu_B,\nu_B\right)
        \nn \\
          &&\hspace{-10.ex}\times 
          S(b,\mu_S,\nu_S)
       J_{f,\text{EEC}}(b,\mu_J,\nu_J)   
       \delta\left(\frac{2|{\bold q}_{T}|}{Q}-\theta\right)\, ,
\eea 
where $U_{tot}\equiv U_BU_HU_JU_S$.

The factorization framework outlined above provides a systematic way to study the EEC in the TMD region and allows for the resummation of large logarithms. It enables the calculation of precise theoretical predictions for the observable in high-energy scattering processes involving hadrons. 
\begin{table}[htb]   

\begin{center}
\begin{tabular}{|c|c|c|c|c|}
\hline   Accuracy & $H$, $J$, $S$, $B$ & $\gamma_{\text{cusp}}$ & $\gamma$ & $\beta$  \\ 
\hline   LL & Tree & 1 loop & - & 1 loop  \\    
\hline   NLL & Tree & 2 loop & 1 loop & 2 loop  \\    
\hline   NNLL & 1 loop & 3 loop & 2 loop & 3 loop  \\    
\hline   N$^3$LL & 2 loop & 4 loop & 3 loop & 4 loop  \\    
\hline   
\end{tabular}
\caption{Classification of the resummation accuracy in terms of the fixed-order expansions of boundary term, anomalous dimensions, and beta function in the TMD region}
\label{table:1}
\end{center}   
\end{table}

Table.~\ref{table:1} lists the ingredients required up to N$^3$LL. The hard function is known at $O(\alpha_s^2)$ in~\cite{Idilbi_2006,Becher:2006mr}. The soft function has been calculated at $O(\alpha_s^2)$ in~\cite{Moult:2018jzp,Li:2016ctv}. The EEC jet function is available up to $O(\alpha_s^3)$ in~\cite{Echevarria_2016,Luo:2019hmp,Luo:2019bmw,Luo:2020epw,Echevarria:2015usa,Ebert:2020qef}. The beam function has been calculated up to $O(\alpha_s^3)$~\cite{Gehrmann:2012ze,Gehrmann_2014,L_bbert_2016,Echevarria_2016,Luo:2019bmw,Luo:2019hmp,Luo:2020epw,Luo:2019szz,Luo_2021}. Finally, the analytic expression for the four loop cusp anomalous dimension, needed to solve the renormalization group evolution equations at N$^3$LL, was obtained recently~\cite{Henn:2019swt,Moult:2022xzt}.

The fact that the factorization involves the universal back-to-back TMD soft function enables us to incorporate hadronization and non-perturbative corrections in a universal framework applicable to conventional TMD observables.

For the soft rapidity anomalous dimension, the implementation of the non-perturbative model is done as in conventional TMD observables at the level of evolution,
\bea
U_{tot}\rightarrow U_{tot}\exp\left(g_K(b)\ln\frac{\nu}{\nu_S}\right)\, .
\eea
Here $g_K(b)=-0.42\ln\left(1+b^2/b^2_{\text{max}}\right)$, is the model function for the non-perturbative component of the rapidity anomalous dimension following the model in~\cite{sun2015universal,Prokudin:2015ysa}. We set $b_{\text{max}}=0.561$.

For the hadronization model of the EEC jet function and TMD beam function, we can assume a generic multiplicative ansatz and the TMD beam function. The 
\begin{align}
    \sqrt{S}J_{f,\text{EEC}}(b,\mu_0,\nu_0)&=\sqrt{S_{\text{pert.}}}J^{\text{OPE}}_{f,\text{EEC}}(b,\mu_0,\nu_0)j_f(b)\,,
    \nn \\
    \sqrt{S}B_{i/p}(x,b;\mu_0,\nu_0)&=\sqrt{S_{\text{pert.}}}B^{\text{OPE}}_{i/p}(b,\mu_0,\nu_0)f_i(b)\,. 
\end{align}
Here $S_{\text{pert.}}$ is the perturbative expression for the soft function, $j_f(b)$ and $f_i(b)$ is the multiplicative ansatz for hadronization effects in the EEC jet function and TMD beam function respectively. The scales $\mu_0$ and $\nu_0$ are arbitrary in the soft, beam, and jet functions. 


Following the model and parameters in~\cite{sun2015universal,Prokudin:2015ysa} we use $f_i(b)=\exp(-0.212b^2)$ and $j_{f}(b)=\exp(-0.59b-0.03b^2)$.
Thus, combining all elements at the level of the cross section we can collect all non-perturbative contributions in a single function,
\bea
\label{eq:totnp}
   F^{\text{NP}}_{i}(b)\equiv j_i(b)f_i(b)\exp\left[-0.42\ln\left(1+\frac{b^2}{b^2_{\text{max}}}\right)\ln\left(\frac{\nu}{\nu_S}\right)\right] .
\eea 

The cross section reads
\bea
\label{eq:TMDfacnp}
    \frac{d\Sigma_N}{dQ^2d\theta}&=&\int dx_B\, x_B^{N}
        H(Q^2,\mu_H)
        \int  d^2{\bold q}_{T}\,  \frac{d^2\bold{b}}{(2\pi)^2} 
        \nn \\
        &&\hspace{-10.ex}\times \exp[-i{\bold q}_T\cdot \bold{b}]U_{tot}F^{\text{NP}}_{f}(b)B^{\text{OPE}}_{f/p}\left(b,x_B,\mu_B,\nu_B\right)
        \nn \\
          &&\hspace{-10.ex}\times 
          S^{\text{pert.}}(b,\mu_S,\nu_S)
       J^{\text{OPE}}_{f,\text{EEC}}(b,\mu_J,\nu_J)   
       \delta\left(\frac{2|{\bold q}_{T}|}{Q}-\theta\right)\, .
\eea 

Further details on hadronization and non-perturbative corrections can be found in~\cite{Li:2021txc} and references therein.

\subsection{Target Fragmentation Region}
\label{subsec:TFR}
We now review the EEC factorization in TFR. 
The detailed derivation of the factorization theorem with SCET in the TFR is given in~\cite{Cao:2023oef}. We note that recently the factorization for the EEC observable in $e^+e^-$ annihilation has been derived within the context of the light-ray operator product expansion (OPE) in~\cite{Chen:2023zzh}, where a similar factorized form as the EEC factorization in the DIS~\cite{Cao:2023oef} was obtained. It will be fascinating to see if the derivation using the light-ray OPE can apply to the EEC case where an external hadronic state is present.   

The expression for $\Sigma_N(Q^2,\theta)$ is given by:
\bea\label{eq:def-eef}
&& \frac{d\Sigma_N}{dQ^2d\theta} 
=  \frac{\alpha^2 }{Q^4}
\int dx_B x_B^{N-1}
L_{\mu\nu} \nn  \\ 
&&   \times
\int 
d^4x
e^{i q \cdot x }
\,  
\langle P |j^{\mu\dagger}(x)    \, 
\hat{{\cal E}}(\theta) \, 
j^\nu  (0) | P \rangle 
\,, 
\eea 
with $L_{\mu\nu}$ the lepton tensor the same as DIS.
The inserted normalized asymptotic energy flow operator $\hat{{\cal E}}(\theta)$ measures the energy deposited in the detector at a given angle $\theta$~\cite{Sveshnikov:1995vi,Tkachov:1995kk,Korchemsky:1999kt,Bauer:2008dt} normalized to the energy $E_P$ of the incoming proton
\bea 
\hat{{\cal E}}(\theta) |X\rangle \equiv 
\sum_{i\in X} \frac{E_i}{E_P} \delta( \theta  - \theta_i ) | X\rangle \,.  
\eea 
The contribution of the energy flow operator in the soft region will be power suppressed by the factor $\frac{E_i}{E_P}$.

We further match the second line in Eq.~(\ref{eq:def-eef}) to the SCET matrix as
\bea 
&& \int 
d^4x
e^{i q \cdot x }
\,  
\langle P |j^{\dagger\mu}(x)    \, 
\hat{{\cal E}}(\theta)\, 
j^\nu  (0) | P \rangle  
= \int 
d^4x 
e^{i q \cdot x } \nn \\ 
&& \times 
\Bigg( C_q^{\mu\nu}(x) 
\langle P |
{\bar \chi}_n(x^-)Y^\dagger(0) \frac{\gamma^+}{2} \hat{{\cal E}}(\theta) Y(0)\chi_n(0)
| P \rangle   \nn \\ 
&& +
 C_g^{\mu\nu}(x) 
\langle P| {\cal B}_\perp (x^-){\cal Y}^\dagger(0)  
\hat{{\cal E}}(\theta) {\cal Y}(0){\cal B}_\perp (0)  |P \rangle 
\Bigg)  \,.
\eea 
which contains only the gauge invariant collinear quark and gluon fields $\chi$ and ${\cal B}_\perp$, respectively~\cite{Stewart:2010qs}.
In addition, we have the soft Wilson lines $Y$ and ${\cal Y}$ in the fundamental and the adjoint representation, respectively. The soft Wilson lines decouple the interaction between the collinear and the soft sectors. Here we note that 
\bea\label{eq:eycommute} 
[\hat{{\cal E}},Y] = [\hat{{\cal E}},{\cal Y}] = 0\,,
\eea 
since $\hat{{\cal E}}(\theta)$ and $Y({\cal Y})$ act on different sectors.
Now we use the identity $Y^\dagger Y  = {\cal Y}^\dagger {\cal Y}= 1$ to reach
\bea 
&& \int 
d^4x
e^{i q \cdot x }
\,  
\langle P |j^{\dagger\mu}(x)    \, 
\hat{{\cal E}}(\theta)\, 
j^\nu  (0) | P \rangle  \nn \\ 
&=& \int 
d^4x 
e^{i q \cdot x } 
\Bigg( C_q^{\mu\nu}(x) 
\langle P |
{\bar \chi}_n(x^-) \frac{\gamma^+}{2} \hat{{\cal E}}(\theta) \chi_n(0)
| P \rangle   \nn \\ 
&& +
 C_g^{\mu\nu}(x) 
\langle P| {\cal B}_\perp (x^-) 
\hat{{\cal E}}(\theta){\cal B}_\perp (0)   |P \rangle 
\Bigg)  \,.
\eea 
We can further derive the hard tensor $C_q^{\mu\nu}$ and $C_g^{\mu\nu}$ is the same as the hard tensor in inclusive DIS by noting that 
\begin{itemize}
\item The above derivation closely follows the SCET derivation of the inclusive DIS cross section in~\cite{Bauer:2002nz}, except for the 
existence of the collinear operator $\hat{{\cal E}}(\theta)$;
\item By substituting the collinear operator $\hat{{\cal E}}(\theta)$ with the identity operator $1=\sum_X|X\rangle \langle X|$, we recover the hadron tensor in the standard inclusive DIS cross section. Meanwhile, Eq.~(\ref{eq:def-eef}) reduce to the inclusive DIS cross section; 
\item The hard coefficients remain unaffected whether using the collinear operator $\hat{{\cal E}}(\theta)$ or the identity operator in the collinear function. This is because the hard coefficients are independent of the details of the collinear sector.
\end{itemize}
Immediately, the factorization of the EEC can be obtained:
\bea\label{eq:fact-x}
\frac{d\Sigma_N}{dQ^2d\theta}
&=&  \sum_{i= q,g} \int dx_B x_B^{N-1}  \nn \\
&& \hspace{-5.ex}
\times \int \frac{dz}{z} f_{\lambda}\hat{\sigma}_{\lambda,i}\left(\frac{x_B}{z},Q^2 \right) f_{i,{\rm EEC}}(z,P^+\theta) \,. 
\eea 
where $f_{i,{\rm EEC}}$ is the quarks NEEC 
\bea\label{eq:fqx} 
&& f_{q,{\rm EEC}}(z,\theta) 
\equiv  \int \frac{dy^-}{4\pi} e^{- i z P^+ \frac{y^-}{2}}  \nn \\ 
&&\hspace{10.ex} \times   \langle P |
{\bar \chi}_n\left(\frac{y^-}{2}n^\mu\right) \frac{\gamma^+}{2} 
\hat{{\cal E}}(\theta)  \chi_n(0)
| P \rangle  \,, 
\eea 
and $f_{g,{\rm EEC}}$ is the gluon NEEC 
\bea\label{eq:fgx}
&& f_{g,{\rm EEC}}(z,\theta)  
= 
\int\frac{dy^-}{4\pi } e^{- i z P^+ \frac{y^-}{2} }   \nn \\ 
&&
\hspace{10.ex} 
\times 
P^+ 
\langle P| {\cal B}_\perp
\left(\frac{y^-}{2}n^\mu \right)   
\hat{{\cal E}}(\theta)   
{\cal B}_\perp (0)  |P \rangle  \,.
\eea 
$\hat{\sigma}_{\lambda,i}$ is the partonic DIS cross section. The corresponding flux is given by 
\bea\label{eq:flux} 
f_{T} = 1-y+\frac{y^2}{2}\,, \quad f_L = 2-2y \,, 
\eea 

We notice that in the TFR, the soft radiations are fully encompassed in the measurement, and therefore the soft modes do not lead to any logarithmic enhancement contributions. This is different from the TMD region measurement, where the soft contribution leads to the enhanced contribution which eventually gives rise to the perturbative Sudakov factor that suppresses the distribution in the TMD region exponentially. 

When $\theta P^+ \gg \Lambda_{\text{QCD}}$, the NEEC can be matched onto the collinear PDFs, with all $\theta$ dependence occurring only in the perturbative matching coefficients. In this way, since $f_{\rm EEC}$ is dimension-less, the $P^+\theta$ will show up in the form of $\ln\frac{P^+\theta}{\mu}$. Therefore, $\frac{d\Sigma_N}{dQ^2d\theta}$ could also be written as
\bea
\label{eq:fact-N}
\frac{d\Sigma_N}{dQ^2d\theta}
&=&  \frac{d\hat{\Sigma}_{T,N}}{dQ^2d\theta}+2\frac{d\hat{\Sigma}_{L,N}}{dQ^2d\theta}
+\frac{Q^4}{2s^2}\frac{d\hat{\Sigma}_{T,N-2}}{dQ^2d\theta}
\nn \\
&& \hspace{-5.ex}-\frac{Q^2}{s}\left(\frac{d\hat{\Sigma}_{T,N-1}}{dQ^2d\theta}+2\frac{d\hat{\Sigma}_{L,N-1}}{dQ^2d\theta}\right)\, ,
\eea 
where we defined
\bea
\label{eq:fact-Nin}
\frac{d\hat{\Sigma}_{\lambda,N}}{dQ^2d\theta}
&=&  \sum_{i= q,g} \int du \, u^{N-1}  \nn \\
&& \hspace{-5.ex}
\times \hat{\sigma}_{\lambda,i}\left(u,Q^2 \right) f_{i,{\rm EEC}}\left(N,\ln \frac{Q\theta}{u\mu}\right) \,, 
\eea 
with $u = \frac{x_B}{z}$ and we have used the fact that $P^+ = \frac{Q}{x_B} = \frac{Q}{z u}$ in the Breit frame. The $\mu$-dependence in other forms through the strong coupling and the collinear PDFs are suppressed in the $f_{i,{\rm EEC}}$, where $f_{i,{\rm EEC}}(N,\ln\frac{Q\theta}{u\mu})$ is the NEEC in the Mellin space, 
\bea\label{eq:mellin-s} 
f_{i,{\rm EEC}}(N,\ln\frac{Q\theta}{u \mu})
= \int_0^1 dz\, z^{N-1} f_{i,{\rm EEC}}(z,\ln\frac{ Q\theta}{ z u \mu})\,. \,\, 
\eea

When $Q\gg \theta Q \gg \Lambda_{\rm QCD}$ NEEC can be matched onto PDF
\bea\label{eq:IN} 
&&f_{i,{\rm EEC}}(N,\ln\frac{Q\theta}{u\mu }) \nn \\ 
&=& f_i(N,\mu) - 
I_{ij}\left(
N, \ln\frac{Q\theta}{u\mu}
\right) f_j(N+1,\mu) \,, 
\eea 
the $I_{ij}$ is a perturbatively calculable matching coefficient and the index $j$ runs over the possible initial parton species in the proton, including the quarks, the anti-quarks, and the gluon.

The NEEC satisfies the modified DGLAP evolution equation
\bea \label{eq:evo2}
&& \frac{d}{d\ln\mu^2} f_{i,{\rm EEC}}(N,\ln\frac{Q\theta}{u \mu}) \nn \\
&=&
\sum_j \int d\xi \xi^{N-1} 
P_{ij}\left(\xi\right)
 f_{j,{\rm EEC}}(N,\ln\frac{Q\theta}{\xi \, u \mu})
 \,, 
\eea 
where $P_{ij}$ is the vacuum splitting function. The solution of this RG equation at NLL level of accuracy is given in~\cite{Cao:2023oef}.
Solving this equation, the NLL NEEC receives the compact analytic form 
\bea\label{eq:fNLLevo} 
&& f_{i,{\rm EEC}}(\mu) = f_i(N,\mu)\nn \\ 
 &&  -    
   {\cal D}_{ik}^N(\mu,\mu_0)
  \, I_{kj}( \ln\frac{Q\theta}{u\mu_0}) f_{j}(N+1,\mu_0)   \nn \\ 
 &&-
 \frac{\alpha_s(\mu_0)}{2\pi}
 {\cal N}_{ik}   [2P_{kj}^{(0)}(N)-2P_{kj}^{(0)}(N+1)]f_j(N+1,\mu_0) \,. \nn \\ 
\eea 
Here $I_{ij}(\ln\frac{Q\theta}{u\mu_0})$ is the NLO matching coefficient in Eq.~(\ref{eq:IN}) evaluated at scale $\mu_0$, and the evolution factor ${\cal D}^N_{ij}(\mu,\mu_0)$ is nothing but the DGLAP evolution in the Mellin space, 
\bea\label{eq:D}
{\cal D}^N_{ij}(\mu,\mu_0) = 
\exp\left[\int_{\mu_0}^{\mu} d\ln\mu^2 P(N,\mu)\right]_{ij} \,.
\eea 
The correction to the DGLAP evolution starts from $\alpha_s^nL^{n-1}$ order, in which  
\bea
{\cal N}_{ij}
&=&  \int_{\mu_0}^\mu d\ln{\mu_1^2}
{\cal D}_{ik}^N(\mu,\mu_1) \tilde{P}_{kl}(N,\mu_1)
{\cal D}_{lj}^N(\mu_1,\mu_0) \,, \quad 
\eea 
Here we have defined 
\bea\label{eq:Ptilde} 
\tilde{P}_{ij} (N)\equiv 
\int dz z^{N-1} P_{ij}(z) \ln z \,.
\eea 
Further details can be found in~\cite{Cao:2023oef} and references therein.

In the limit of extremely small angles, we anticipate the $d\Sigma_N/d\theta$ pattern indicates the presence of a free hadron phase where the energy is uniformly distributed. In this phase, the energy deposit within the region bounded by the polar angle being less than $\theta$ is proportional to $\theta^2$. As NEEC is proportional to the distribution of energy with respect to the polar angle, We expect
\bea \label{eq:np} 
\frac{d\Sigma_N}{d\theta}|_{\text{NP}}\propto \theta\, .
\eea 
The analogous pattern has also been observed in the final state jet through the utilization of CMS open data~\cite{Komiske:2022enw}.

\section{Numerical Results}
\label{sec:num}
In this section, we explore the EEC distributions with two distinct collision energies. The interaction of 18 GeV electrons with 275 GeV protons at the EIC with $\sqrt{s}=140.7$ GeV, and the interaction of 22 GeV electrons with 2 GeV protons at CEBAF with $\sqrt{s}=13.3$ GeV. 

For the EIC kinematics, we set the parameters as $N=4$ and $Q=20$ GeV, while for the CEBAF kinematics, we consider $N=4$ and $Q=3$ GeV. For all the numerical results, we use the {\tt PDF4LHC15\_nnlo\_mc} PDF sets~\cite{Butterworth:2015oua} with the associated strong coupling provided by {\tt LHAPDF6}~\cite{Buckley:2014ana}.

\begin{figure}[htbp]
  \begin{center}
   \includegraphics[scale=0.9]{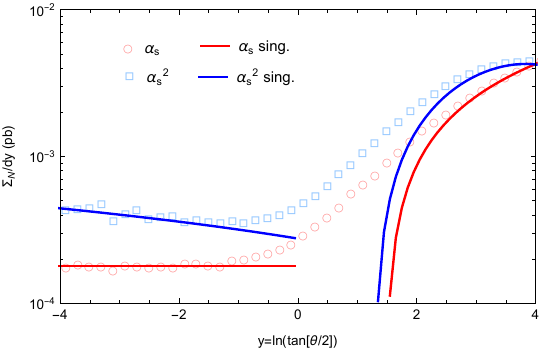} 
    \caption{Comparison between the $\ln\theta$ leading singular contributions
with the full fixed-order calculations in both the forward and backward regions.}
  \label{fg:sing}
 \end{center}
\end{figure}

Firstly, we validate the factorization formalism by comparing the leading singular $\ln\theta$ contributions predicted by the factorization theorem with the complete $\alpha_s$ and $\alpha_s^2$ calculations of the distribution $d\Sigma_N/dy$, where $y\equiv\ln\tan \frac{\theta}{2}$.
As $\theta$ is small or large, the $\ln \theta$ terms dominate the distribution, and the singular contribution should coincide with the full fixed-order calculation.
The comparison is shown in Fig.~\ref{fg:sing} utilizing EIC kinematics where the renormalization and factorization scales are set to be Q. The full fixed-order calculations are obtained numerically using {\tt nlojet++}~\cite{Nagy:2005gn}. Remarkably, in both the small $y$ and large $y$ regions, we observe excellent agreement between the leading singular terms predicted by the factorization formula and the full fixed-order calculations. This comparison validates the factorization theorem.  In the forward region when $y>-1$ and the backward region when $y<2.5$ Fig.~\ref{fg:sing} starts to show differences between the leading singular and full QCD calculations, corresponding to transition regions between resummation and fixed-order calculations.


In the TMD and the TFR regions, the logarithmic enhancements can spoil the convergence of the perturbative expansion. Therefore, the resummation of these logarithms to all orders in the strong coupling is necessary for reliable predictions to compare with experimental data.

In the TMD region, the resummed cross section can be evaluated by evolving the hard, soft, beam, and jet functions in Eq.~(\ref{eq:TMDfac}) from their canonical scales to common rapidity and renormalization scales, $\nu$ and $\mu$ respectively. Here we choose the canonical resummation scales as
\bea
\label{eq:scales}
\mu=\mu_H=\nu=\nu_J=\nu_B=Q, \quad \mu_J=\mu_S=\mu_B=\frac{2e^{-\gamma_E}}{b^*}\,.
\eea
To avoid the Landau pole at large $b$, we employ a local $b^*$ prescription~\cite{Sun:2014dqm,Prokudin:2015ysa} freezing out the virtuality scales. Specifically, we have
\bea
b^*=\frac{b_T}{\sqrt{1+b_T^2/b^2_{\text{max}}}} \,,\qquad\frac{2e^{-\gamma_E}}{b_{\text{max}}}=2\text{GeV}\,.
\eea
We choose $2e^{-\gamma_E}/b_{\text{max}}=2\text{GeV}$. This ensures that the scale used in the PDFs is larger than 1 GeV when we vary the scale by a factor of two.

\begin{figure}[htbp]
    \begin{minipage}{\linewidth}
    \includegraphics[scale=0.9]{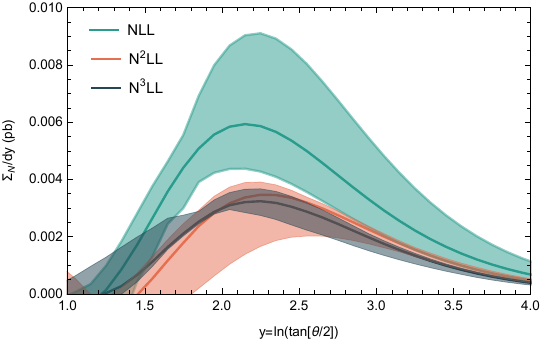}
    \end{minipage}
    \begin{minipage}{\linewidth}
    \includegraphics[scale=0.9]{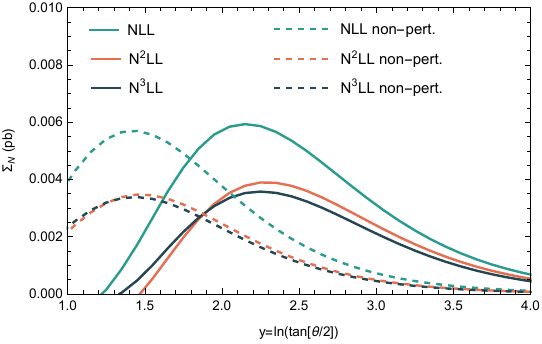}
    \end{minipage}
    \caption{Resummed y distributions for EEC in the TMD region.}
    \label{fg:error}
\end{figure}

Fig.~\ref{fg:error} presents the resummed distributions in the TMD region using EIC kinematics. The upper panel shows the result without non-perturbative models, the scale uncertainties are evaluated by varying scales up and down in Eq.~(\ref{eq:scales}) by a factor of 2 independently. We observe large corrections from NLL to N$^2$LL and a good perturbative convergence from N$^2$LL to N$^3$LL. Furthermore, we find that the scale uncertainties are significantly reduced for the N$^3$LL compared to the lower accuracy distributions. For the non-perturbative models discussed in section \ref{subsec:BR} the result is present in the lower panel of Fig.~\ref{fg:error}. The non-perturbative corrections shift the peak of the cross sections to smaller $y$. The non-perturbative effects presented here are consistent with those reported in~\cite{Li:2021txc}.

In the TFR, we consistently choose $\mu_h=\mu$, allowing us to evaluate the resummed cross section by evolving the NEEC in Eq.~(\ref{eq:fact-N}) from $\mu_0$ to $\mu$. In this case, we select the canonical resummation scales as follows:
\bea
\label{eq:scalesTFR}
\mu=Q, \qquad \mu_0=\frac{Q\theta}{2}\,.
\eea
The scale uncertainties are evaluated by varying scales in Eq.~(\ref{eq:scalesTFR}) up and down by a factor of 2 independently. 

\begin{figure}[htbp]
    \begin{minipage}{\linewidth}
    \includegraphics[scale=0.9]{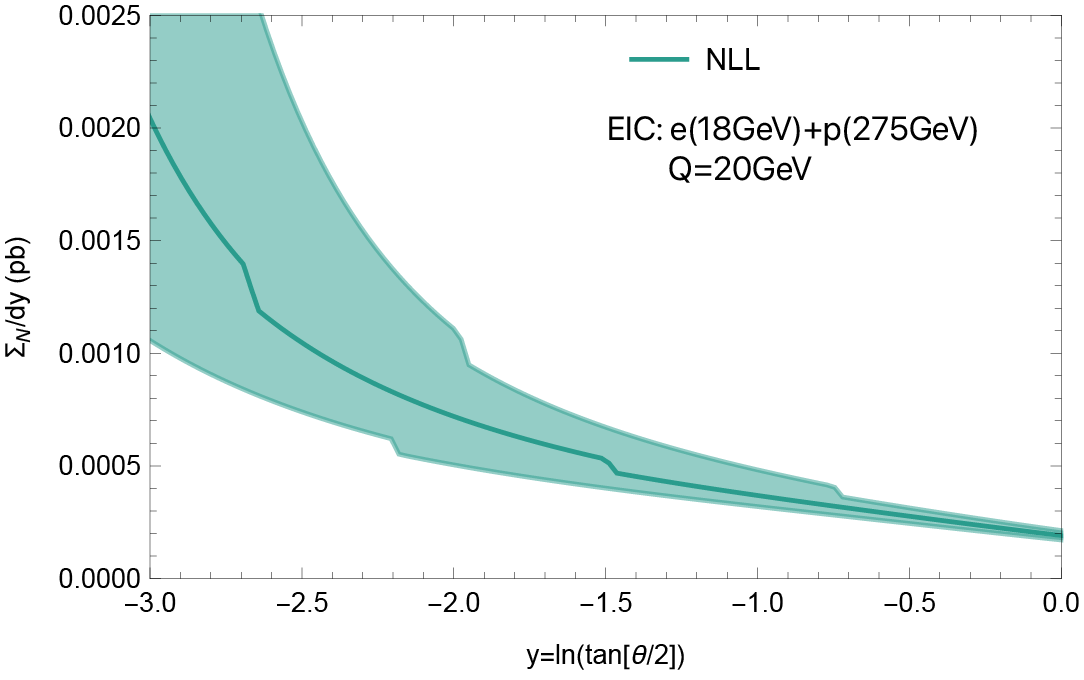}
\end{minipage}
    \begin{minipage}{\linewidth}
    \includegraphics[scale=0.9]{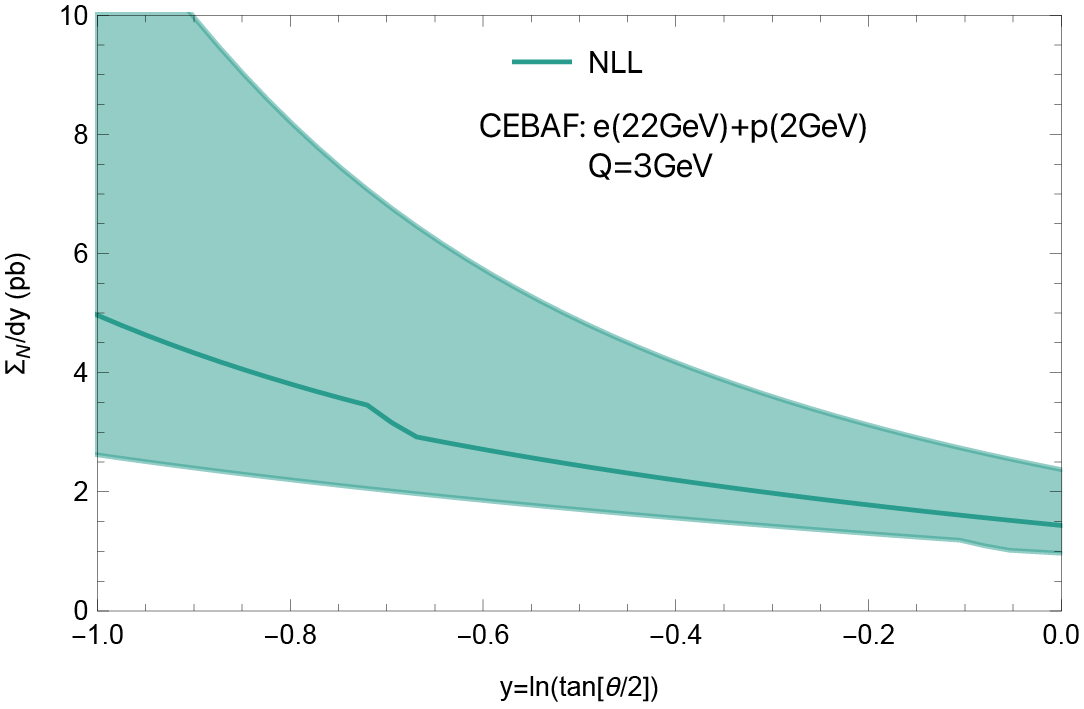}
    \end{minipage}
    \caption{Resummed $y$ distributions for EEC within the TFR, depicted in the upper and lower panels for EIC kinematics and CEBAF kinematics, respectively.}
    \label{fg:errorTFR}
\end{figure}

The upper panel of Fig.~\ref{fg:errorTFR} illustrates the resummed distributions in the TFR region for EIC kinematics. 
When $y<-3$, $\mu_0$ is comparable with $\Lambda_{\text{QCD}}$. The perturbative calculation is no longer valid in this regime. The non-smoothness observed in the curve is a consequence of the impact of quark masses. As $\mu_0$ crosses the threshold of a quark mass, it necessitates a modification in the number of quark flavors, resulting discontinuous in the R.H.S of Eq.~(\ref{eq:fNLLevo}).

The lower panel of Fig.~\ref{fg:errorTFR} demonstrates the resummed distributions in the TFR region in CEBAF kinematics. Similarly, When $y < -1$, $\mu_0$ is comparable with $\Lambda_{\text{QCD}}$. Comparing these results to the fixed-order calculations depicted in Fig.~\ref{fg:sing}, we can see that the resummation effects play a significant role in the small angle region. This resummation enhances the distribution several times compared to the $\alpha^2_s$ calculation for $y$ around $-3$. Furthermore, it is worth emphasizing that the distribution at small angles exhibits no suppression since there is no conventional Sudakov factor in the NEEC. This unique property of the NEEC stands in marked contrast to the behavior observed in TMD PDFs, which experience exponential suppression in the small transverse momentum
region induced by the Sudakov factor.

\begin{figure}[htbp]
    \begin{minipage}{\linewidth}
    \includegraphics[scale=0.9]{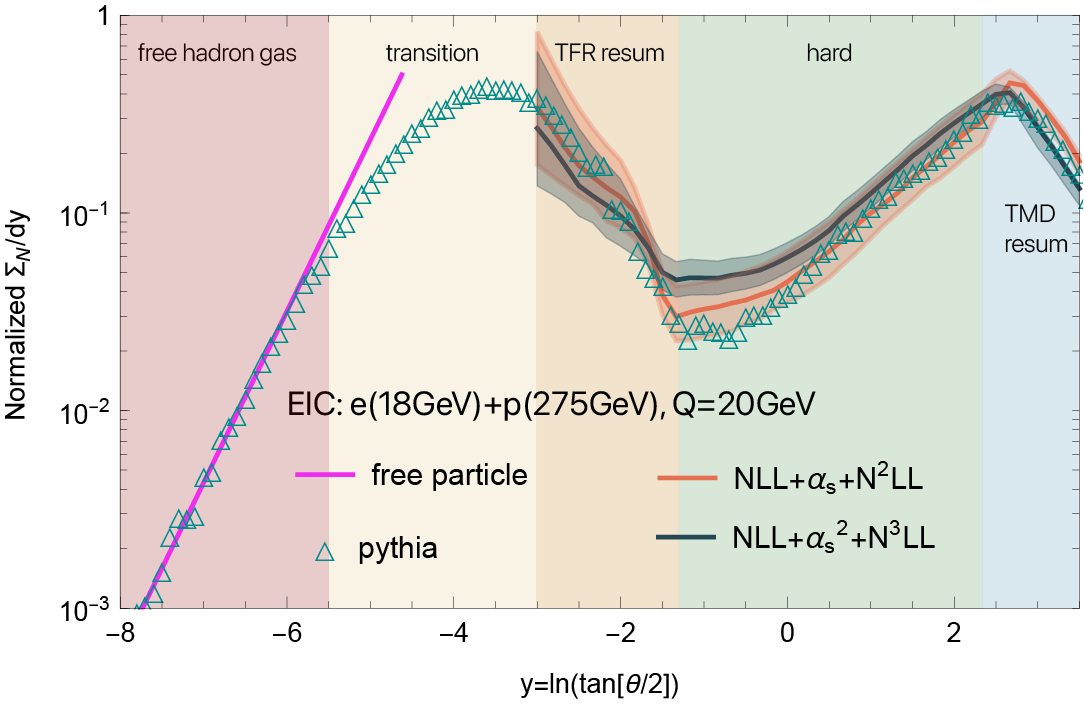}
    \end{minipage}
    \begin{minipage}{\linewidth}
    \includegraphics[scale=0.9]{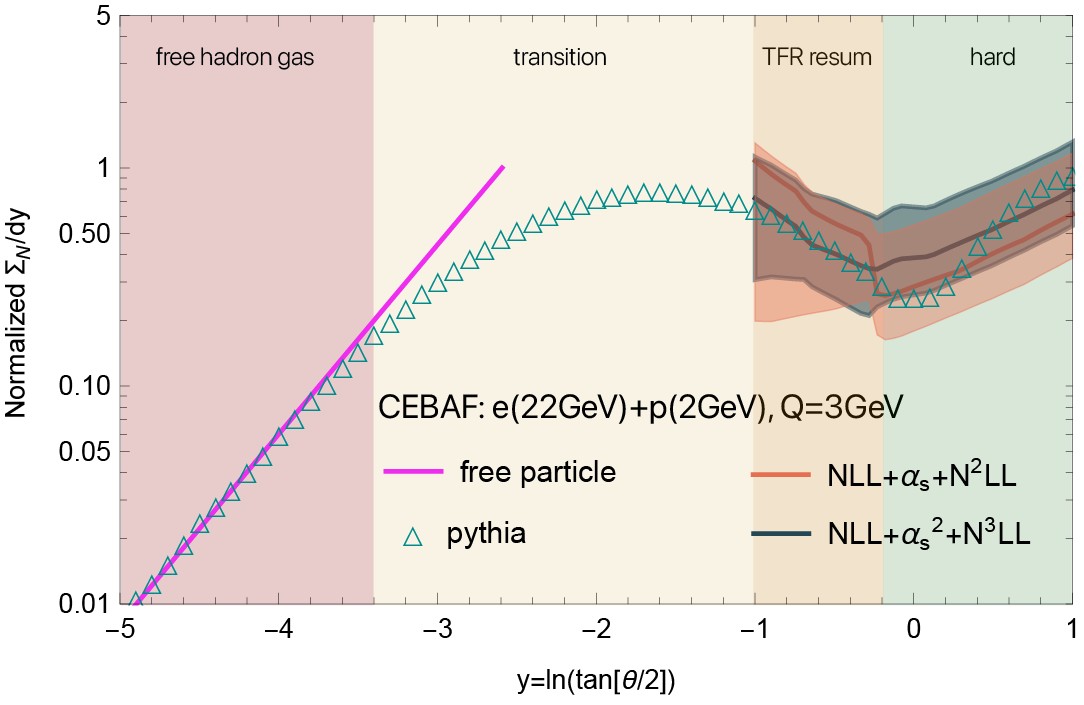}
    \end{minipage}
    \caption{Comparison of EEC between the SCET predictions without non-perturbative effects, free hadron gas model, and PYTHIA simulations running without hadronization modeling. The upper panel displays the results for EIC kinematics, while the lower panel showcases the results for CEBAF kinematics.}
    \label{fg:pythia}
\end{figure}

The final distributions without non-perturbative effects for EIC and CEBAF are presented in the upper and lower panels of Fig.~\ref{fg:pythia}, respectively. In the TMD region, we match the N$^2$LL (N$^3$LL) resummed distributions to the QCD LO (NLO) ones. In the TFR region, we match the NLL resummed distributions to the QCD LO (NLO) ones.
The uncertainties in the full spectrum are evaluated by varying all scales up and down independently by a factor of two. We compare our calculations to {\tt PYTHIA}~\cite{Sjostrand:2014zea, Bierlich:2022pfr} simulations without a hadronization modeling.

In EIC kinematics, the distribution is normalized to the central curve over the range $-3<y<3.5$. The distributions are described by the fixed-order results for $-1<y<2.5$, by the sum of resummed and non-singular power corrections for $y>3$ and $-3<y<-2$. In the region $-2<y<-1.3$ and $2.3<y<3$, we apply our matching scheme where the cross-section smoothly transitions from the resummed to the fixed-order cross-section. The matching scheme is defined as
\bea
\label{eq:match}
\frac{d\Sigma_N}{dQ^2dy}&=&(1-f^2)\frac{d\Sigma_N}{dQ^2dy}\Bigg|_{\text{QCD}}
\nn\\
&&\hspace{-10.ex}+f^2\left(\frac{d\Sigma_N}{dQ^2dy}\Bigg|_{\text{non-sing.}}+\frac{d\Sigma_N}{dQ^2dy}\Bigg|_{\text{res}}\right)
\eea
where
\bea
\label{eq:scheme}
f=
\left\{
\begin{aligned}\frac{1}{2}\left[\cos\left(\frac{\cos\theta-a}{a-b}\pi\right)+1\right]&&(2.3<y<3)
\\
\frac{1}{2}\left[\cos\left(\frac{\cos\theta-c}{c-d}\pi\right)+1\right]&&(-2<y<-1.3)
\end{aligned}\right.\,.
\eea
Here, $a$, $b$, $c$ and $d$ equal to $\cos\theta$ with $\theta$ associated with $y = 3$, $y=2.3$, $y=-2$ and $y=-1.3$, respectively. 
A similar matching procedure and detailed discussion about matching can be found in ~\cite{Becher_2019}.

In CEBAF kinematics, the distribution is normalized to the central curve over the range  $-1<y<1$. For $-0.2<y<1$ the distributions are described by the fixed-order results, and for $y<-0.3$ by the sum of TFR resummed and non-singular power corrections. In region $-0.2<y<-0.1$ we impose the same matching scheme as Eq.~(\ref{eq:match}), but with
\bea
\label{eq:scheme2}
f=
\frac{1}{2}\left[\cos\left(\frac{\cos\theta-a}{a-b}\pi\right)+1\right]\,,
\eea
where $a$ and $b$ equal to $\cos\theta$ with $\theta$ associated with $y=-0.2$ and $y=-0.1$, respectively.

In the perturbative region, the matching result agrees reasonably well with the partonic {\tt PYTHIA} simulation. The NLL+$\alpha_s$+N$^2$LL agrees better with {\tt PYTHIA}.
There is a difference between NLL+$\alpha^2_s$+N$^3$LL and NLL+$\alpha_s$+N$^2$LL mainly due to the $\mathcal{O}(\alpha^2_s)$ corrections. 


In the extreme forward region, when $y<-5.5$ in EIC kinematics and $y<-3.4$ in CEBAF kinematics, we fit the un-normalized {\tt PYTHIA} distribution with the non-perturbative model $a_{\text{NP}}\theta$ to observe the free hadron gas phase. Even without hadronization, We observe a nearly perfect $d\Sigma_N/dy\propto\theta^2$ scaling, as expected above in Eq.~(\ref{eq:np}), corresponding to uniformly distributed partons. 

Furthermore, in Fig.~\ref{fg:pythia}, we observe a distinct phase transition. The transition from the TFR resummation region to the free hadron gas region, connected by a non-perturbative transition region, occurs at approximately $\theta\sim0.1$ rad in EIC kinematics and $\theta\sim0.7$ rad in CEBAF kinematics.
When comparing the EIC kinematics distribution to the CEBAF kinematics distribution, we observe that the free hadron gas region and the transition region shift to larger angles in the CEBAF kinematics distribution. This shift is expected since the transition occurs as $\theta \sim O(\Lambda_{\text{QCD}}/Q)$. Consequently, CLAS holds the potential for probing NEEC in the non-perturbative region, which essentially enables direct imaging of the confining transition to free hadrons.

\begin{figure}[htbp]
    \begin{minipage}{\linewidth}
    \includegraphics[scale=0.9]{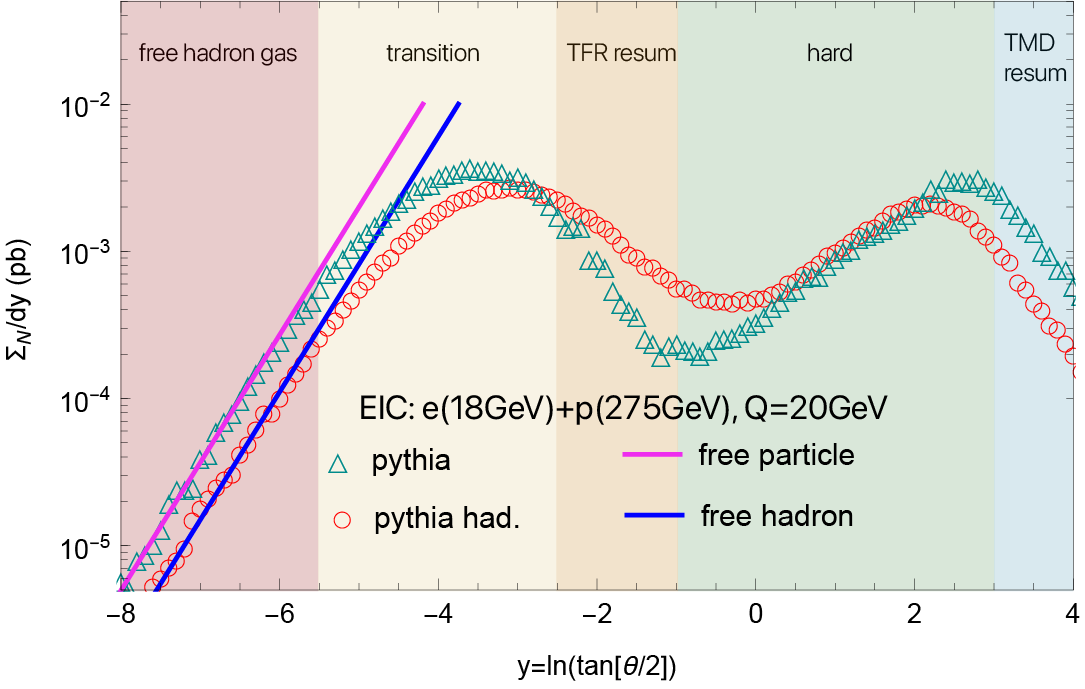}
    \end{minipage}
    \begin{minipage}{\linewidth}
    \includegraphics[scale=0.9]{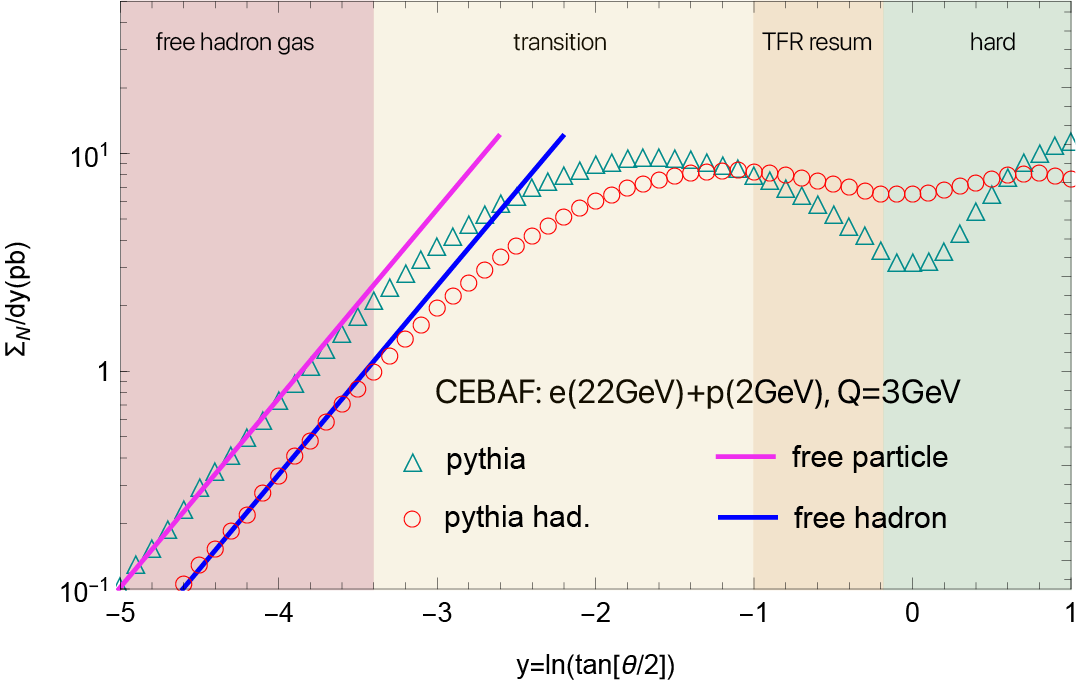}
    \end{minipage}
    \caption{Comparison of EEC between {\tt PYTHIA} simulations with and without hadronization. The upper panel displays the results for EIC kinematics, while the lower panel showcases the results for CEBAF kinematics.}
    \label{fg:pythiah2}
\end{figure}

\begin{figure}[htbp]
    \includegraphics[scale=0.9]{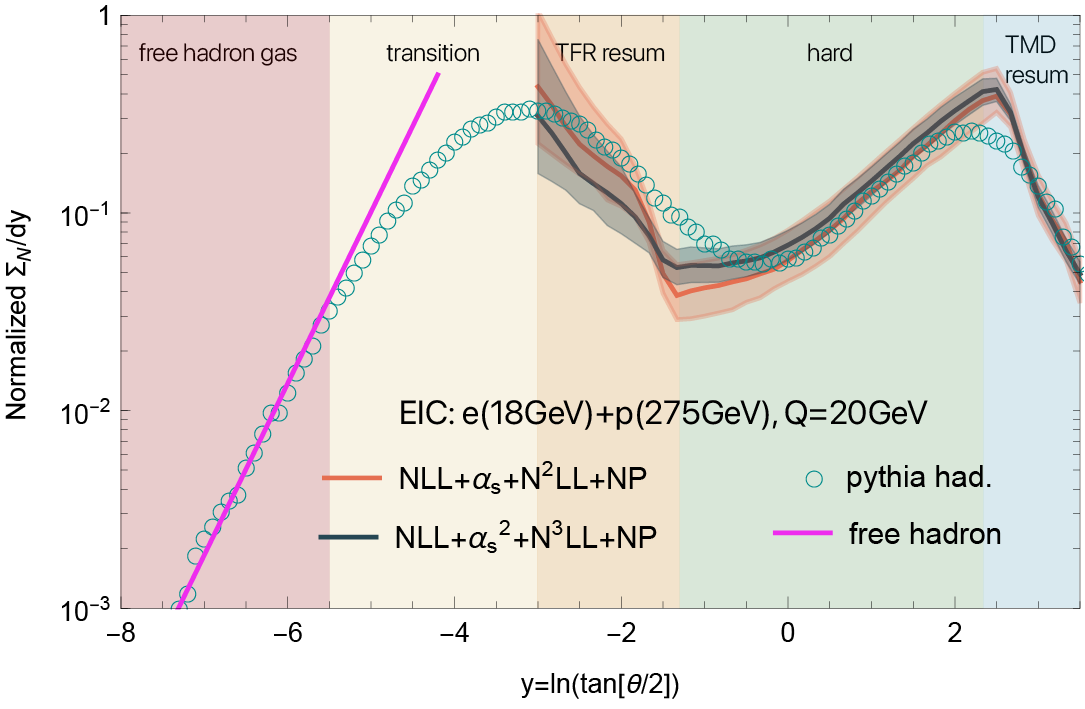}
    \caption{The comparison of EEC between the SCET predictions with hadronic {\tt PYTHIA} simulations for EIC kinematics.}
    \label{fg:pythiah}
\end{figure}

In Fig.\ref{fg:pythiah2}, we compare the simulated {\tt PYTHIA} result with and without hadronization for both EIC and CEBAF kinematics. 
We observe that for $y<-5.5$ in EIC kinematics and $y<-3.4$ in CEBAF kinematics, the $d\Sigma_N/dy\propto\theta^2$ scaling persists, indicating the presence of uniformly distributed hadrons.
By comparing the distributions in Fig.\ref{fg:pythiah}, we can see that the inclusion of hadronization effects, as implemented in {\tt PYTHIA}, enhances the distribution in the central region while the distribution in the free hadron gas, transition, and TMD region is suppressed.

The final distributions with non-perturbative effects for EIC are presented in Fig.~\ref{fg:pythiah}. We use the same matching strategy as discussed above. 
We compare our calculations with {\tt PYTHIA} simulations that include hadronization. We include the non-perturbative model in the TMD resummed results while the TFR resummed results are unchanged. The comparison in the TMD resummed region can be used to validate the universality of the non-perturbative model and parameters introduced in Eq.~(\ref{eq:totnp}) extracted from SIDIS. 


\section{Conclusion}
\label{sec:con}
In this work, we explore the Bjorken $x$ weighted EEC in DIS from the TFR to CFR. In both regions, a factorization theorem can be derived with SCET, based on which the logarithms can be resummed to all orders in $\alpha_s$. 
The singular distributions can be derived from the factorized formula, which are compared against the full fixed-order QCD calculations up to NLO. This comparison serves two main purposes. Firstly, the numerical agreement observed in the TMD region and TFR validates our factorization formalism. Secondly, the point at which the distributions deviate indicates the region where power corrections become significant. Additionally, we present the resummation results up to NLL in the TFR and N$^3$LL in the TMD region. 

In the extremely small angle limit, the free hadron gas model is introduced to investigate the non-perturbative distribution. We compared our predictions to partonic {\tt PYTHIA} simulations. Between the hadron gas phase region and the perturbative resummation region, a transition phase is observed. We note that the transition region from perturbative parton phase to non-perturbative region for CEBAF begins at $\theta\sim0.7$ rad, indicating CLAS may have a good opportunity to probe the non-perturbative NEEC.

The non-perturbative and hadronization effects in the TMD region were investigated by considering non-perturbative form factors extracted from the semi-inclusive hadron production in DIS. Incorporating these non-perturbative models, we also presented the comparison of our predictions to {\tt PYTHIA} simulations.


The recent progress in understanding EEC in DIS holds great promise, and we firmly believe that it will play a pivotal role in advancing our comprehension of nucleon structure in the years to come.

\begin{acknowledgments}
We thank Xiaohui Liu for the useful discussions and careful reading of the draft. This work is supported by the Natural Science Foundation of China under contract No.~12175016 (H.~C. and Z.~M.) and No.~12275156 (H.~T.~L.).
\end{acknowledgments}

\begin{widetext} 
\appendix 
\section{Useful Identities}
When we calculate the singular $\ln\theta$ contributions predicted by the factorization theorem in Eq.~(\ref{eq:TMDfac}), all $b$ dependence in the soft, jet, and beam function will show up in the form of $\ln b$. Therefore $\mathbf{b}$ can be integrated analytically using the following formula.
The Fourier Transformation between $\bold{b}_T$ and $\bold{q}_T$ can be derived from
\begin{align}
    \int \frac{d^2 \bold{q}_T}{2\pi} \exp[i \bold{q}_T \cdot \bold{b}_T] \frac{1}{\mu^2} \left( \frac{\mu^2}{q_T^2}\right)^{1+\alpha} = -\frac{e^{-2\alpha \gamma_E}}{2\alpha} \frac{\Gamma(1-\alpha)}{\Gamma(1+\alpha)} \left( \frac{b_T^2 \mu^2}{4 e^{-2 \gamma_E}} \right)^{\alpha}.
\end{align}
The explicit transformations of the logarithmic terms from $b_T$ space to $q_T$ space are 
\begin{align} 
    1 \to& (2\pi)\delta^{(2)}(\bold{q}_T)
    \nonumber \\
    \ln\frac{b^2\mu^2}{4 e^{-2 \gamma_E}} \to &  -\frac{2}{ q_T^2}
    \nonumber \\
    \ln^2\frac{b^2\mu^2}{4 e^{-2 \gamma_E}} \to&  -4\frac{1}{ q^2_T} \ln \frac{\mu^2}{q^2_T}
    \nonumber \\
    \ln^3\frac{b^2\mu^2}{4 e^{-2 \gamma_E}} \to& -6\frac{1}{  q_T^2} \ln^2 \left(\frac{\mu^2}{q_T^2}\right)-4\zeta(3)(2\pi)\delta^{(2)}(\bold{q}_T) 
    \nonumber \\
    \ln^4\frac{b^2\mu^2}{4 e^{-2 \gamma_E}} \to& -8\frac{1}{q_T^2}\ln^3 \left(\frac{\mu^2}{q_T^2}\right)
   +32 \zeta (3)\frac{1}{ q_T^2}\, 
\end{align}

\section{Renormalization group evolution in TMD factorization}

The RG evolution equation for the hard, beam, jet, and soft functions is given by
\bea
\label{eq:HRG}
\mu \frac{d}{d\mu} H(Q^2,\mu) &=& \gamma_H \>H(Q^2,\mu)\, ,\nn \\
\mu \frac{d}{d\mu} J_{EEC}(z,b,\mu,\nu) &=& \gamma_J(\mu,\nu) J_{EEC}(z,b,\mu,\nu)\, , \nn \\
\nu \frac{d}{d\nu} J_{EEC}(z,b,\mu,\nu) &=& -\frac{1}{2}\gamma_\nu(b,\mu) J_{EEC}(z,b,\mu,\nu)\, , \nn \\
\mu \frac{d}{d\mu}B_q(x,b,\mu,\nu)&=& \gamma_B(\mu,\nu) B_q(x,b,\mu,\nu)\, , \\
\nu \frac{d}{d\nu} B_q(x,b,\mu,\nu) &=& -\frac{1}{2}\gamma_\nu(b,\mu) B_q(x,b,\mu,\nu)\, ,\nn  \\
\mu\frac{d}{d\mu}S_{EEC}(b,\mu,\nu) &=& \gamma_S(\mu,\nu)\> S_{EEC}(b,\mu,\nu), \nn \\ 
\nu\frac{d}{d\nu}S_{EEC}(b,\mu,\nu) &=& \gamma_\nu (b,\mu)\> S_{EEC}(b,\mu,\nu)\,, \nn
\eea
where the anomalous dimensions $\gamma_H$ is given by
\bea
\label{eq:pertexpanomdim}
\gamma_H(\mu) &=& 2C_F\gamma_{\rm cusp}\ln\frac{Q^2}{\mu^2}+2\gamma_q\, ,\nn \\
\gamma_B(\mu,\nu) &=& C_F \gamma_{\text{cusp}}\ln\frac{\nu^2}{Q^2} + 2\gamma_B^q\, , \nn \\
\gamma_J(\mu,\nu) &=& C_F \gamma_{\text{cusp}}\ln\frac{\nu^2}{Q^2} + 2\gamma_J^q\,,   \\
\gamma_S(\mu,\nu)&=&2C_F\gamma_{\rm cusp}\ln\frac{\mu^2}{\nu^2}-2\gamma_{EEC}^s\,,\nn  \\
\gamma_\nu(b,\mu)&=&-4\int_{\mu_0}^{\mu}\frac{d\mu'}{\mu'}C_F\gamma_{\text{cusp}}+2\gamma_{EEC}^r(\mu_0)\,.  \nn 
\eea
The solution to the RG equations for the hard function in Eqs.~(\ref{eq:HRG}) has the form in Eq.~(\ref{eq:hardfuncevol}), with the hard function evolution factor has the form
\bea
\label{eq:UH}
U_H(Q^2,\mu,\mu_H) &=& \exp \Big[  4 C_FS(\mu,\mu_H)-2A_H (\mu,\mu_H)  \Big ]\Big ( \frac{\mu_H^2}{Q^2} \Big )^{2C_FA(\mu,\mu_H)}, 
\eea
where the functions $S, A_{\text{cusp}}$ and $A_H$ are defined as
\bea
\label{eq:Smufmui_Amufmui_AHmufmui}
 S(\mu_f,\mu_i) &=& -\int_{\alpha_s(\mu_i)}^{\alpha_s(\mu_f)} \frac{d\alpha}{\beta[\alpha]}\gamma_{\text{cusp}}[\alpha] \int_{\alpha_s(\mu_i)}^\alpha \frac{d\alpha '}{\beta[\alpha ']},\nn \\
A(\mu_f,\mu_i) &=& -\int_{\alpha_s(\mu_i)}^{\alpha_s(\mu_f)}\frac{d\alpha}{\beta[\alpha]}\gamma_{\text{cusp}}[\alpha],  \\
A_H(\mu_f,\mu_i) &=& -\int_{\alpha_s(\mu_i)}^{\alpha_s(\mu_f)}\frac{d\alpha}{\beta[\alpha]}\gamma^{q}[\alpha], \nn
\eea
We define $r\equiv\frac{\alpha_s(\mu_f)}{\alpha_s(\mu_i)}$
The perturbative expansion of $S(\mu_f,\mu_i)$ needed for N$^3$LL resummation is given by
\bea
\label{eq:SpertexpN3LL}
S(\mu_f,\mu_i) &=& \frac{\gamma_0^{\text{cusp}}}{4\beta_0^2}\Bigg \{ \frac{4\pi}{\alpha_s(\mu_i)} \Big ( 1-\frac{1}{r}-\ln r\Big ) + \Big ( \frac{\gamma_1^{\text{cusp}}}{\gamma_0^{\text{cusp}}}-\frac{\beta_1}{\beta_0}\Big )(1-r+\ln r )+\frac{\beta_1}{2\beta_0}\ln^2r\nn \\
&+& \frac{\alpha_s(\mu_i)}{4\pi}\Bigg [\Big ( \frac{\beta_1\gamma_1}{\beta_0\gamma_0^{\text{cusp}}} - \frac{\beta_2}{\beta_0} \Big )(1-r+r\ln r) +\Big ( \frac{\beta_1^2}{\beta_0^2}-\frac{\beta_2}{\beta_0}\Big )(1-r)\ln r  \nn\\
&-&\Big(\frac{\beta_1^2}{\beta_0^2}-\frac{\beta_2}{\beta_0}-\frac{\beta_1 \gamma_1^{\text{cusp}}}{\beta_0\gamma_0^{\text{cusp}}}+\frac{\gamma_2^{\text{cusp}}}{\gamma_0^{\text{cusp}}}\Big)\frac{(1-r)^2}{2}\Bigg ] \\
&+&\left [\frac{\alpha_s(\mu_i)}{4\pi}\right ]^2\Bigg [
                                       \left  (\frac{\beta_1\beta_2}{\beta_0^2} - \frac{\beta_1^3}{2\beta_0^3}- \frac{\beta_3}{2\beta_0} +\left (\frac{\gamma_2^{\text{cusp}}}{\gamma_0^{\text{cusp}}}-\frac{\beta_2}{\beta_0}+\frac{\beta_1^2}{\beta_0^2}-\frac{\beta_1\gamma_1^{\rm cusp}}{\beta_0\gamma_0^{\rm cusp}}\right)\frac{\beta_1r^2}{2\beta_0} \right )\ln r \nn \\
                                        &+& \left (\frac{\gamma_3^{\rm cusp}}{\gamma_0^{\rm cusp}}-\frac{\beta_3}{\beta_0} + \frac{2\beta_1\beta_2}{\beta_0^2}+\frac{\beta_1^2}{\beta_0^2}\left (\frac{\gamma_1^{\rm cusp}}{\gamma_0^{\rm cusp}} - \frac{\beta_1}{\beta_0}\right )-\frac{\beta_2\gamma_1^{\rm cusp}}{\beta_0\gamma_0^{\rm cusp}}
                                        -\frac{\beta_1\gamma_2^{\rm cusp}}{\beta_0\gamma_0^{\rm cusp}}
                                        \right )\frac{(1-r)^2}{3} \nn \\
                                        &+&\left (\frac{3\beta_3}{4\beta_0}-\frac{\gamma_3^{\rm cusp}}{2\gamma_0^{\rm cusp}}
                                       +\frac{\beta_1^3}{\beta_0^3}-\frac{3\beta_1^2\gamma_1^{\rm cusp}}
                                       {4\beta_0^2\gamma_0^{\rm cusp}}
                                        +\frac{\beta_2\gamma_1^{\rm cusp}}{\beta_0\gamma_0^{\rm cusp}}
                                        +\frac{\beta_1\gamma_2^{\rm cusp}}{4\beta_0\gamma_0^{\rm cusp}}
                                        -\frac{7 \beta_1\beta_2}{4\beta_0^2}\right )(1-r)^2\nn \\
                                        &+&\left (\frac{\beta_1\beta_2}{\beta_0^2}-\frac{\beta_3}{\beta_0}
                                        -\frac{\beta_1^2\gamma_1^{\rm cusp}}{\beta_0^2\gamma_0^{\rm cusp}}
                                        +\frac{\beta_1\gamma_2^{\rm cusp}}{\beta_0\gamma_0^{\rm cusp}}
                                         \right)\frac{1-r}{2} \nn
                                      \Bigg ]
\Bigg \}\, .  
\eea
The corresponding perturbative expansion for $A_{\text{cusp}}(\mu_f,\mu_i)$ is given by
\bea
\label{Aevo}
A_{\text{cusp}}(\mu_f,\mu_i)&=& \frac{\gamma_0^{\text{cusp}}}{2\beta_0}\Bigg\{
\log r + \,
\frac{\alpha_s(\mu_i)}{4\pi}(1-r)\left(\frac{\gamma_1^{\text{cusp}}}{\gamma_0^{\text{cusp}}}-\frac{\beta_1}{\beta_0} \right)\nn \\
&&
+ \left[\frac{\alpha_s(\mu_i)}{4\pi}\right]^2
\Bigg[\frac{\gamma_2^{\text{cusp}}}{\gamma_0^{\text{cusp}}}-\frac{\beta_2}{\beta_0}
-\frac{\beta_1}{\beta_0}\,
\left(\frac{\gamma_1^{\text{cusp}}}{\gamma_0^{\text{cusp}}}-\frac{\beta_1}{\beta_0} \right)
\Bigg]\frac{r^2-1}{2}\nn \\
&&
+ \frac{1}{3}\left[\frac{\alpha_s(\mu_i)}{4\pi}\right]^3\Bigg [\frac{\gamma_3^{\rm cusp}}{\gamma_0^{\rm cusp}}-\frac{\beta_3}{\beta_0} + \frac{\gamma_1^{\rm cusp}}{\gamma_0^{\rm cusp}}\left (\frac{\beta_1^2}{\beta_0^2}-\frac{\beta_2}{\beta_0} \right )  \\
&&
-\frac{\beta_1}{\beta_0}\left (\frac{\beta_1^2}{\beta_0^2}-\frac{2\beta_2}{\beta_0} +\frac{\gamma_2^{\rm cusp}}{\gamma_0^{\rm cusp}}\right )\Bigg ]\left (r^3-1\right )^3
\Bigg \}\,. \nn 
\eea
Finally, the corresponding expansion for $A_H(\mu_f,\mu_i)$ is given by
\bea
\label{AHevo}
A_H(\mu_f,\mu_i)&=& \frac{\gamma_0^{q}}{2\beta_0}\Bigg\{
\log r + \,
\frac{\alpha_s(\mu_i)}{4\pi}\left(\frac{\gamma_1^{q}}{\gamma_0^{q}}-\frac{\beta_1}{\beta_0} \right)
 \nn \\
&&
+ \left[\frac{\alpha_s(\mu_i)}{4\pi}\right]^2
\Bigg[\frac{\gamma_2^{q}}{\gamma_0^{q}}-\frac{\beta_2}{\beta_0}
-\frac{\beta_1}{\beta_0}\,
\left(\frac{\gamma_1^{q}}{\gamma_0^{q}}-\frac{\beta_1}{\beta_0} \right)
\Bigg]\frac{r^2-1}{2}
\Bigg \}\, . 
\eea
Solving the other RG equations in Eq.~(\ref{eq:HRG}) gives the beam, jet, and soft functions evolved to any arbitrary scale, $\mu$, from their values at their natural scales $\mu_B, \mu_J,$ and $\mu_S$, respectively, where large logarithms in their perturbative expansions are minimized. They have the general form given in Eq.~(\ref{eq:BJSevolPos}), 
where the $U_B,U_J,$ and $U_S$ denote the RG evolution factors have the form
\bea
\label{eq:UBJS}
U_B(\mu,\mu_B,\nu,\nu_B) &=& \exp[-2A_{B}(\mu,\mu_B)]\left(\frac{Q^2}{\nu_B^2}\right)^{C_FA(\mu,\mu_J)}\left(\frac{\nu^2}{\nu_B^2}\right)^{-C_FA(\mu,\mu_0)-\gamma^r_{EEC}(\alpha(\mu_0))/2},\nn \\
U_J(\mu,\mu_J,\nu,\nu_J) &=& \exp[-2A_{J}(\mu,\mu_J)]\left(\frac{Q^2}{\nu_J^2}\right)^{C_FA(\mu,\mu_J)}\left(\frac{\nu^2}{\nu_J^2}\right)^{-C_FA(\mu,\mu_0)-\gamma^r_{EEC}(\alpha(\mu_0))/2},  \\
U_S(\mu,\mu_S,\nu,\nu_S) &=& \exp\left[-4C_FS(\mu,\mu_S)+2A_S(\mu,\mu_S)\right]
        \left(\frac{\mu^2_S}{\nu^2_S}\right)^{-2C_FA(\mu,\mu_S)}\left(\frac{\nu^2}{\nu_S^2}\right)^{2C_FA(\mu,\mu_0)+\gamma^r_{EEC}(\alpha(\mu_0))} \nn \,,
\eea
where the function $S$ is defined in Eq.~(\ref{eq:Smufmui_Amufmui_AHmufmui}) and its perturbative expansion needed for N$^3$LL resummation is given in Eq.~(\ref{eq:SpertexpN3LL}). The functions $A_B, A_J,$ and $A_S$ are defined as
\bea
\label{eq:ABAJAS}
A_B(\mu_f,\mu_i)&=& -\int_{\alpha_s(\mu_i)}^{\alpha_s(\mu_f)}\frac{d\alpha}{\beta[\alpha]}\gamma^{q}_B[\alpha],\nn \\
A_J(\mu_f,\mu_i)&=& -\int_{\alpha_s(\mu_i)}^{\alpha_s(\mu_f)}\frac{d\alpha}{\beta[\alpha]}\gamma^{q}_J[\alpha], \\
A_S(\mu,\mu_S) &=& -\int_{\alpha_s(\mu_i)}^{\alpha_s(\mu_f)}\frac{d\alpha}{\beta[\alpha]}\gamma^s_{EEC}[\alpha]\,.\nn
\eea
The corresponding expressions for the perturbative expansions of $A_B, A_J,$ and $A_S$ in Eq.~(\ref{eq:ABAJAS}) needed for N$^3$LL resummation can be obtained by replacing $\gamma^{q}_{0,1,2}$ with $\gamma^{q}_{B,{0,1,2}}, \gamma^{q}_{J,{0,1,2}},$ and $\gamma^{s}_{EEC,{0,1,2}}$, respectively, in Eq.~(\ref{AHevo}).

\end{widetext} 

\bibliographystyle{h-physrev}   
\bibliography{refs}

\end{document}